\documentclass[10pt,letterpaper]{article}
\usepackage{opex3}
\usepackage{amsmath}
\usepackage{color}
\usepackage{cite}

\begin{document}

\title{Squeezed light from a diamond-turned monolithic cavity}

\author{A.\ Brieussel,$^{1,2,*}$ Y.\ Shen,$^{1,3}$ G.\ Campbell,$^{1}$ G.\ Guccione,$^{1}$ J.\ Janousek,$^{1}$ B.\ Hage,$^{4}$ B.\ C.\ Buchler,$^{1}$ N.\ Treps,$^{2}$ C.\ Fabre,$^{2}$ F.\ Z.\ Fang,$^{3}$ X.\ Y.\ Li,$^{3}$ \linebreak[2] T.\ Symul,$^{1}$ and P.\ K.\ Lam$^{1}$
}

\address{$^1$Centre for Quantum Computation and Communication Technology, Department of Quantum Science, The Australian National University, 38a Science Road, Acton, ACT, 2601 Australia\\
$^2$Laboratoire Kastler Brossel, Universit\'e Pierre et Marie Curie, Case 74, 75252 Paris Cedex 05, France\\
$^3$College of Precision Instrument and Opto-electronics Engineering, Key Laboratory of Optoelectronics Information Technology of Ministry of Education, Tianjin University, Tianjin, 300072, China\\
$^4$Universit\"at Rostock, Universit\"atsplatz 3, 18051 Rostock, Germany}

\email{Ping.Lam@anu.edu.au} 

\begin{abstract}
For some crystalline materials, a regime can be found where continuous ductile cutting is feasible.
Using precision diamond turning, such materials can be cut into complex optical components with high surface quality and form accuracy. In this work we use diamond-turning to machine a monolithic, square-shaped, doubly-resonant $LiNbO_3$ cavity with two flat and two convex facets. When additional mild polishing is implemented, the Q-factor of the resonator is found to be limited only by the material absorption loss. We show how our monolithic square resonator may be operated as an optical parametric oscillator that is evanescently coupled to free-space beams via birefringent prisms. The prism arrangement allows for independent and large tuning of the fundamental and second harmonic coupling rates. We measure $2.6\pm0.5$~dB of vacuum squeezing at 1064~nm using our system. Potential improvements to obtain higher degrees of squeezing are discussed.
\end{abstract}

\ocis{(230.4320) Nonlinear optical devices; (130.3730)   Lithium niobate; (310.2790)   Guided waves; (220.1920) Diamond machining; (270.6570) Squeezed states.}

\bibliographystyle{osajnl}

\section{Introduction}

Technologies such as quantum communication, computation, and metrology often require the use of non-classical quantum states such as squeezed states. Over the past two decades, the generation of these states has been optimized, with squeezing of up to 12.7 dB  \cite{PhysRevLett.104.251102} below the shot noise being reported. The most common approach is to use an optical cavity to enhance the non-linear interaction between a bright pump field and a sub-harmonic field at half its frequency via a $\chi^{(2)}$ non-linear crystal. The most common design for such a cavity is the bow-tie configuration \cite{takeno2007observation}. This design is excellent from a research perspective since it allows easy replacement of the nonlinear crystal and is relatively simple to build and align. From the perspective of building the best possible squeezed light source, however, this approach has limitations. The cavity is typically long ($\sim$0.5~m roundtrip) which means active length-stabilization is crucial to counteract acoustic and thermal perturbations to the cavity length. The bow-tie geometry is also not suitable for miniaturization as the larger angles of reflection required would lead to an astigmatic cavity mode. There is also a minimum of six internal surfaces, four mirrors and two crystal faces, that can lead to scattering loss, limiting the escape-efficiency of the cavity.

One approach to solve these issues is to use a monolithic design where the surfaces of the nonlinear crystal are used as cavity mirrors \cite{kurz1992squeezing}. The crystal can be polished and coated to give a specified mode and cavity finesse allowing a compact geometry and potentially lower internal loss. This comes at the cost of tunability since the mode and finesse of the nonlinear cavity cannot be varied. A semi-monolithic design \cite{PhysRevLett.104.251102} has one polished and coated crystal surface combined with an external input/output coupling mirror. This design allows some adjustability but without all the advantages of true monolithic design.
Rather than coating a mirror onto the nonlinear crystal, a cavity can also be formed within the crystal by total internal reflection (TIR). Input and output coupling are then accomplished by frustrated TIR. Such TIR monolithic cavities have previously been investigated for wavelength conversion \cite{schiller1993quadruply,fiedler1993highly,paschotta1994nonlinear} and proposed for the production of quantum light \cite{paschotta1994nonlinear}. The first successful demonstrations, however, have only recently been performed using a whispering gallery resonator to generate bright twin-beam squeezing \cite{PhysRevLett.106.113901} or single photons \cite{PhysRevA.91.023812}.

A key technological challenge is the shaping of the crystal surfaces to form a resonant mode that is as close as possible to the TEM$_{00}$ mode, thereby allowing the output light to couple easily to other optical components. In this paper we introduce a cavity that has been prepared using a precision diamond-tooled lathe and present results demonstrating $2.6\pm0.5$~dB of vacuum squeezing. The design contrasts with the whispering gallery resonator approach in that it features four distinct reflecting surfaces to define a resonant mode, as opposed to a continuous guiding of the light. This approach shares many of the advantages of the whispering gallery design, such as stability and a simple fabrication process, while avoiding some of the difficulties, such as a complex mode structure and the need for high-index coupling prisms \cite{Schunk:14,matsko2009practical}. In particular, the modes in the resonator are defined by the conventional TEM basis which simplifies the coupling to free space modes. We further demonstrate that the coupling rate of both the pump field and squeezed field to free space can be independently tuned.

\begin{figure}
\begin{centering}
\includegraphics[width=0.6\columnwidth]{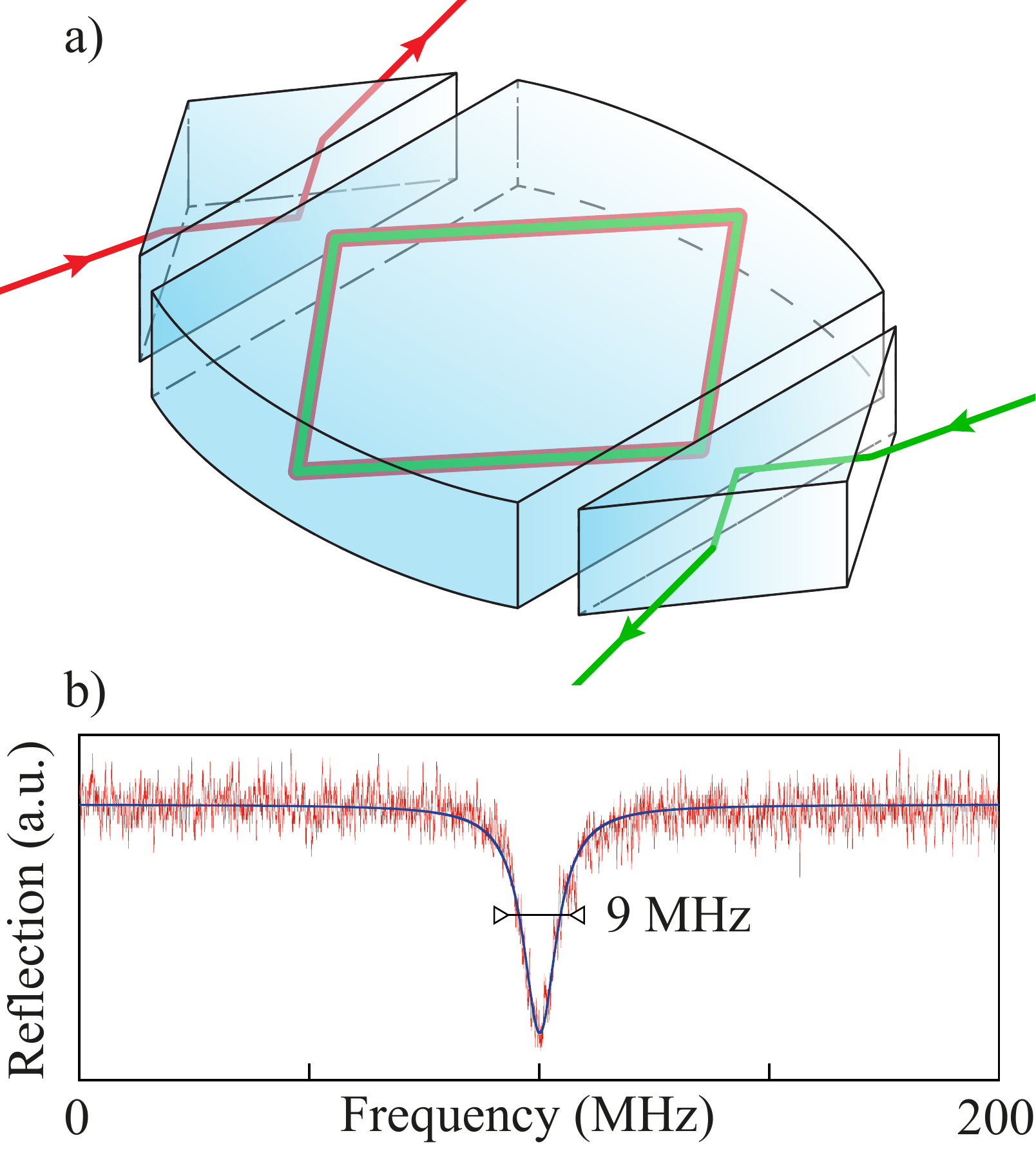}
\par\end{centering}
\caption{(a) Schematic of the resonator. An optical crystal is cut into a square-shaped monolithic resonator that uses total internal reflection to define an optical mode. Two opposite faces are spheroidal to confine the optical mode while the remaining two faces are used to evanescently couple the circulating mode to a free-space optical mode via a prism. (b) Minimum linewidth achievable for the sub-harmonic. The measurement is performed by scanning the voltage applied across the resonator. To observe the intrinsic linewidth, the coupling prism is placed sufficiently far away that coupling losses are negligible relative to losses from absorption and scattering.}
\label{3DSqOPO}
\end{figure}

\section{Resonator design and fabrication}

The machining process provides a simple fabrication process for creating relatively complex monolithic resonators. The shape of the resonator is produced in a single continuous cut rather than distinct grinding steps. This allows flexibility in the design of the resonator, enabling compact sizes and the use of aspheric reflecting surfaces. The curved surfaces, for example, may be cut in the shape of an oblate spheroid to ensure non-astigmatic modes in free space when the resonator is prism-coupled. The curvatures can also be selected to optimize the effective non-linear interaction between the pump and sub-harmonic although, for the resonator presented in this work, larger curvatures were used as a compromise to facilitate mode-matching.

The resonator design, shown conceptually in Fig.~\ref{3DSqOPO}(a), consists of four smooth surfaces arranged in a square. Total internal reflection from these surfaces defines a closed optical path, also in the shape of a square, rotated 45$^\circ$ with respect to the square of the resonator. Two of the reflecting surfaces are curved to form a stable optical cavity. The remaining two surfaces are nearly flat to facilitate prism coupling, although they have a slight convex curvature to ensure that the prism can contact the resonator at the location of the mode waist on the coupling surfaces.

\begin{figure}
\begin{centering}
\includegraphics[width=0.7\columnwidth]{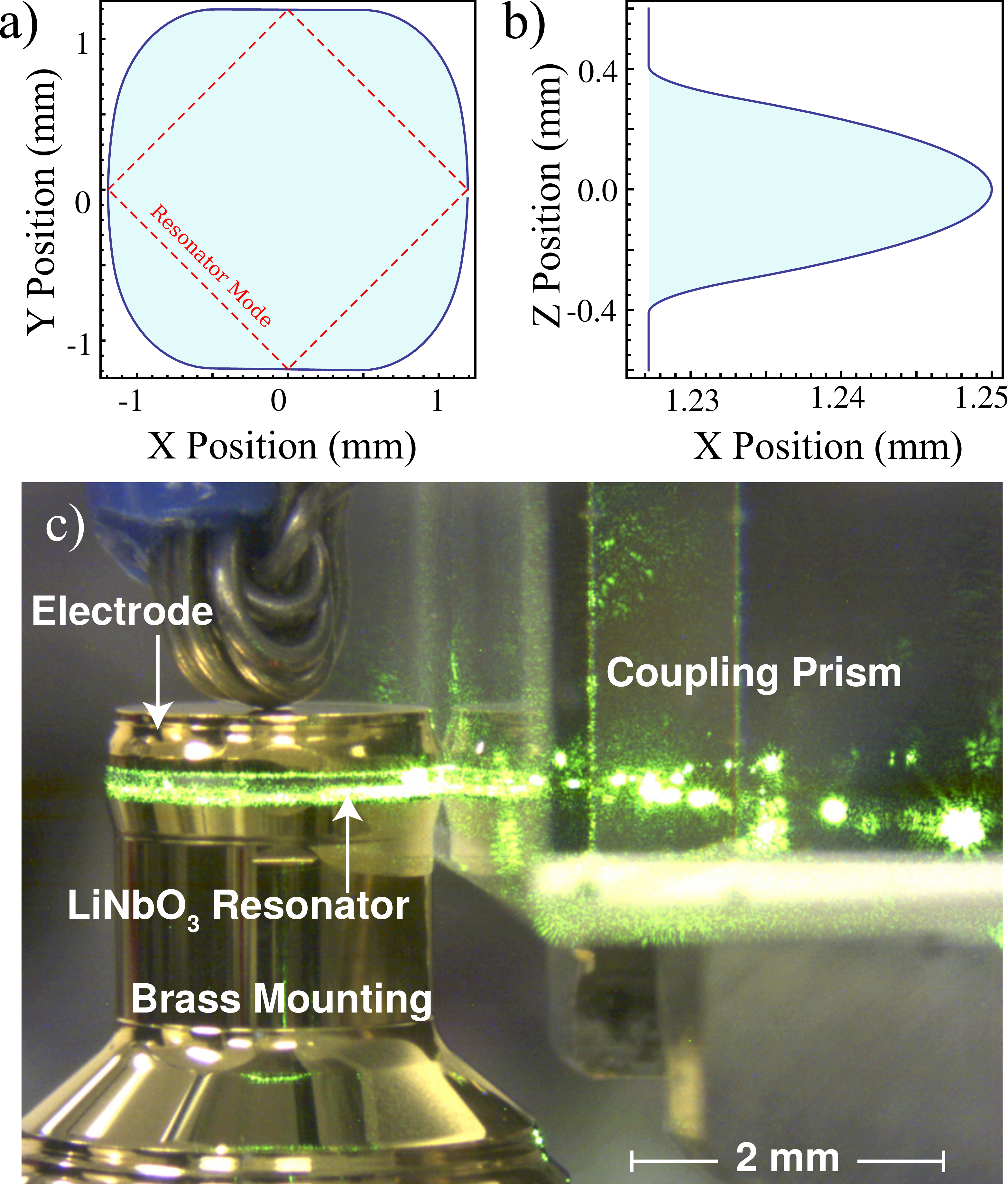}
\par\end{centering}
\caption{Illustration of the shape of the resonator from a top-down perspective (a) and of the curvature of one of the total internal reflection mirrors as seen from the side (b) with an aspect ratio of 40:1. An image of a complete, prism-coupled resonator is shown in (c). The lithium niobate resonator is sandwiched between two brass electrodes, and is illuminated with scattering from 532 nm pump light that is circulating in the resonator. A lead for voltage-tuning the resonator is in contact with the top electrode and a green calcite coupling prism can be seen contacting the resonator from the right.}
\label{fig:couplingpicture}
\end{figure}

To machine this cavity, we start with a 200 $\mu$m thick magnesium oxide-doped lithium niobate (LN) wafer that has been cut into a disk with the extraordinary optical axis oriented normal to the plane of the disk. This material can be phase-matched for second-harmonic generation and optical parametric oscillation between 1064~nm and 532~nm by temperature tuning and is also well suited to diamond cutting. The disk is sandwiched between two brass electrodes that will later facilitate electro-optic tuning and also provide stability during machining. The cutting process is similar to that which is used to fabricate whispering gallery resonators \cite{grudinin2006ultra}, however, the cut depth is a function of the azimuthal angle during cutting so that a non-cylindrically symmetric shape is produced. The corners of the resonator are smoothed to reduce the departure from cylindrical symmetry and simplify the turning process. The resonator has a width of 2.5 mm with mirror curvatures of 3 mm for the curved reflective surfaces, and 22 mm for the coupling surfaces. Figures~\ref{fig:couplingpicture}(a) and \ref{fig:couplingpicture}(b) show the shape of the resonator from top-down and side-on perspectives.

\begin{figure}
\begin{centering}
\includegraphics[width=0.9\columnwidth]{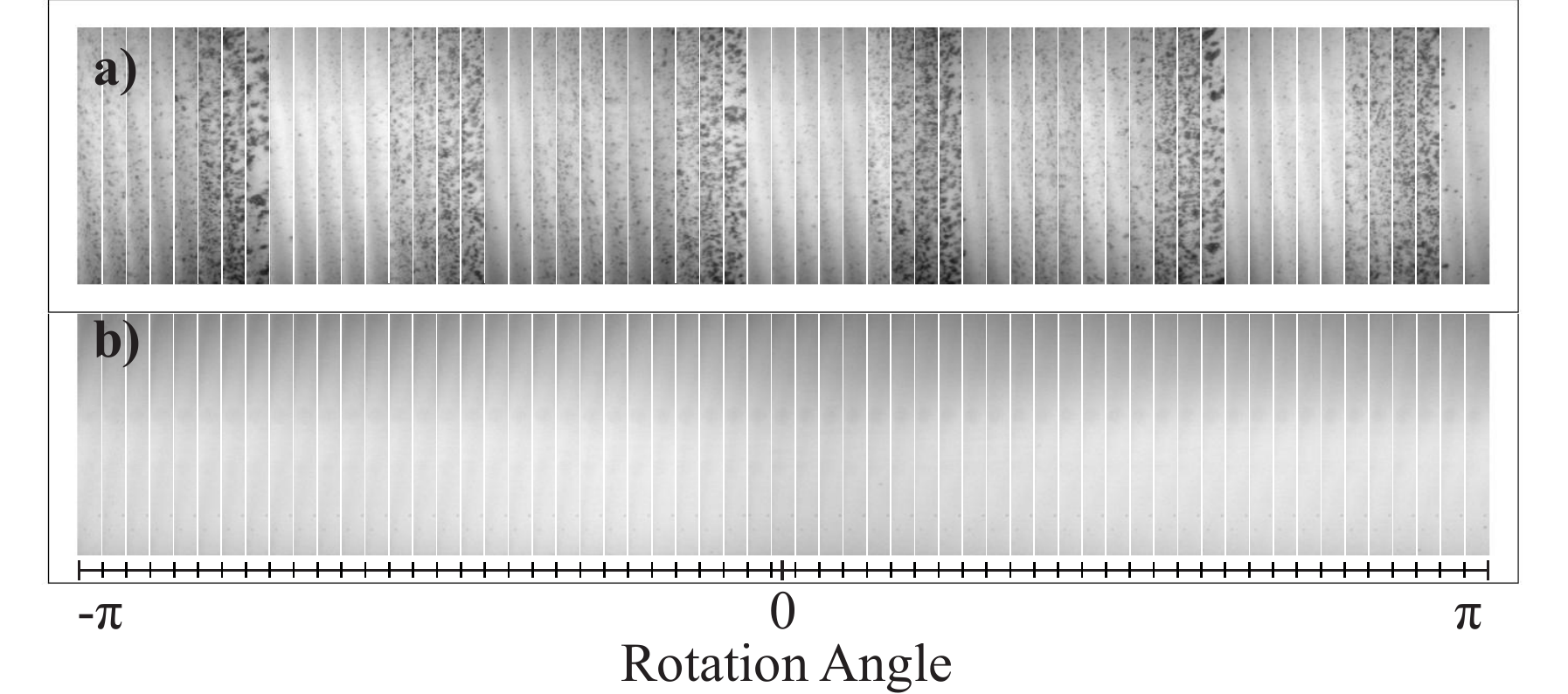}
\par\end{centering}
\caption{Examples of the surfaces that result from (a) brittle and (b) ductile modes of material removal while diamond turning LiNbO$_3$. Each images are the composition of 59 images each taken at one specific angle.}
\label{CutFig}
\end{figure}

Care is taken during the diamond turning process to ensure that the material removal process is ductile, as opposed to brittle, to avoid chipping the surface \cite{ngoi2000ductile,fang1998ductile}. This is accomplished through the use of a diamond tool with a chamfered edge and a very small depth-of-cut for finishing cuts on the resonator. Figure~\ref{CutFig} shows microscope images of the surface of an LN disk taken after roughing and finishing cuts to illustrate the transition from brittle to ductile removal. The ductile machining process means that only a minimum amount of material needs to be removed by polishing, enabling hand-polishing of the four reflective surfaces to optical quality \cite{fuchs1992diamond,grudinin2006ultra}. The polishing process results in a sufficiently smooth surface that the linewidth of the resonators is limited by bulk absorption in the LN rather than by scattering from surface imperfections. Figure~\ref{3DSqOPO}(b) shows an under-coupled cavity resonance for 532 nm light with a linewidth of $9\pm1$~MHz. This is consistent with the bulk-absorption limited linewidth of LN whispering gallery resonators \cite{furst2010naturally} and corresponds to a Q factor of $3.1\times 10^7$.

The inclusion of brass electrodes placed around the resonator provides voltage tuning of the cavity resonances. The electro-optic coefficients for the extraordinarily polarized 532 nm pump field provide a tuning rate of 40 MHz/V while the ordinarily polarized 1064 nm sub-harmonic is tuned at a rate of 6 MHz/V. This tuning is useful for mode-matching to the cavity modes and for characterization of the cavity; however, an applied voltage enhances photo-refractive effects in LN through increased photo-conductivity \cite{bryan1985magnesium,volk1994optical} and limits the use of voltage tuning to situations where the intra-cavity power of the pump field is very small. Figure~\ref{fig:couplingpicture}(c) shows a prism-coupled resonator with tuning electrodes and an attached high-voltage lead.

\section{Prism coupling}

The cavity modes are coupled to free-space modes by prism coupling \cite{matsko2009practical,schiller1992fused,fiedler1993highly,pan2003highly}. A prism is brought into the evanescent field of one of the total internal reflections. Provided that the prism has a sufficiently high refractive index to frustrate the total internal reflection, light will be coupled out of the resonator to a propagating free-space optical mode. The large radius of curvature of the coupling surfaces ensures that the mode will be uniformly coupled out of the resonator and thus the modes both inside and outside the resonator will be close to TEM modes.

\begin{figure}
\begin{centering}
\includegraphics[width=1\columnwidth]{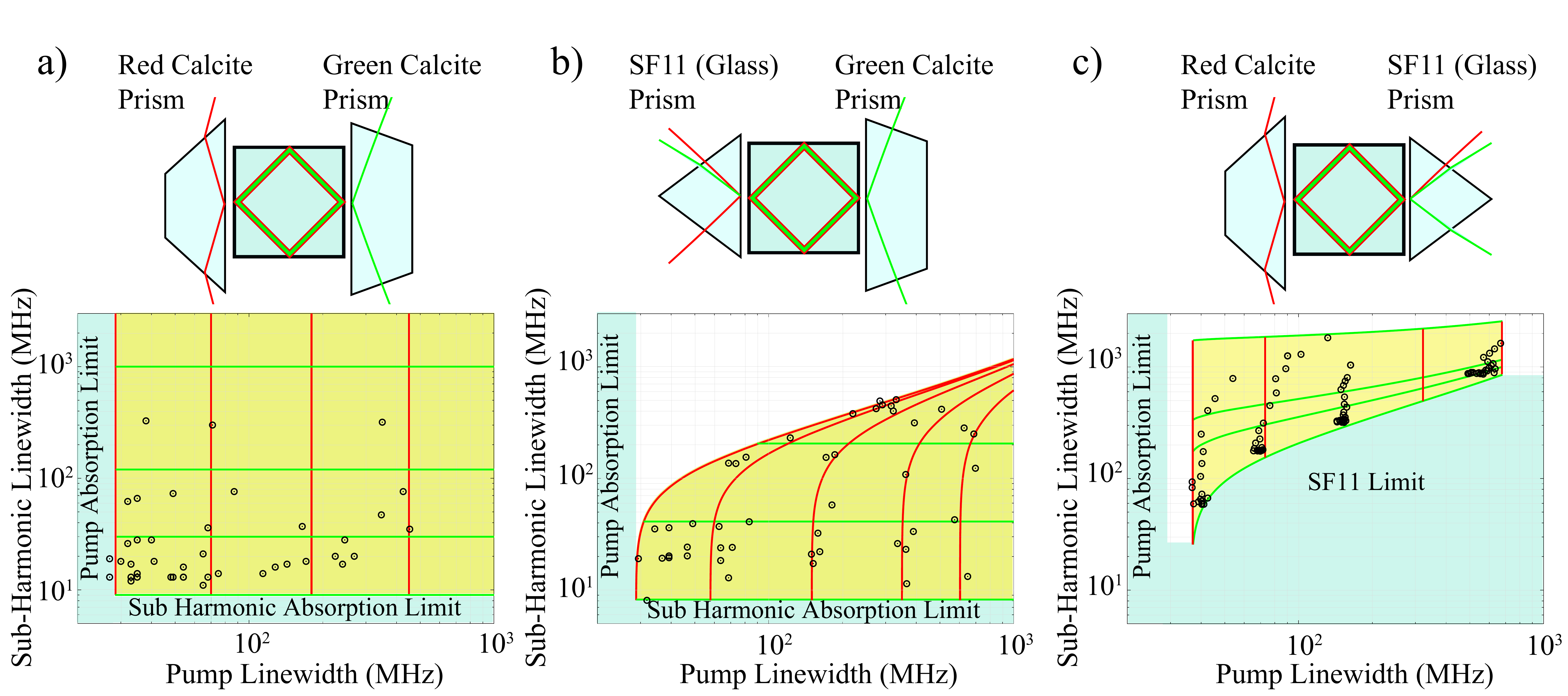}
\par\end{centering}
\caption{\label{fig:FinesseGreenRedThreePrism} (Top) Illustration of the configuration of the prism couplers. (Bottom) Linewidths attainable by the system for each configuration. The yellow areas represent the possible finesses for 1064~nm sub-harmonic and 532~nm pump. The dark circles illustrate experimental data. The red lines are found using a lossless model and correspond to a displacement of the sub-harmonic coupler. The green lines correspond to the lossless model for displacements of the pump coupler.}
\end{figure}

The square shape of the resonator results in an internal angle of incidence of 45$^\circ$ for each of the total internal reflections. This large angle affords significant flexibility in the choice of coupling prisms because it reduces the refractive index that is required to frustrate total internal reflection. Similar coupling methods for lithium niobate whispering gallery resonators restrict the choices for prism materials to options such as diamond \cite{furst2010naturally} or TiO$_2$.

We choose a calcite prism with its optic axis oriented in the horizontal plane, making an angle of 40$^\circ$ with the normal of the coupling face, to couple the pump light into the resonator. The birefringence allows the selective coupling of only the ordinarily polarised pump mode (at 532~nm) while the extraordinarily polarized sub-harmonic (at 1064~nm) still undergoes total internal reflection \cite{fiedler1993highly}. We refer to this prism as the green calcite prism. Similarly, the sub-harmonic can be selectively coupled using a calcite prism that has its optic axis oriented vertically so that the ordinarily polarized pump mode undergoes total internal reflection. We refer to this prism as the red calcite prism. To reduce losses in the squeezed field, the red calcite prism has its end faces cut such that the out-coupled light is incident at Brewster's angle. The geometries of these prisms and their orientation relative to the resonator is illustrated in Fig.~\ref{fig:FinesseGreenRedThreePrism}(a). For alignment purposes, a glass (SF11) prism can be used to couple both the pump and sub-harmonic.

\begin{figure}
\begin{centering}
\includegraphics[width=0.7\columnwidth]{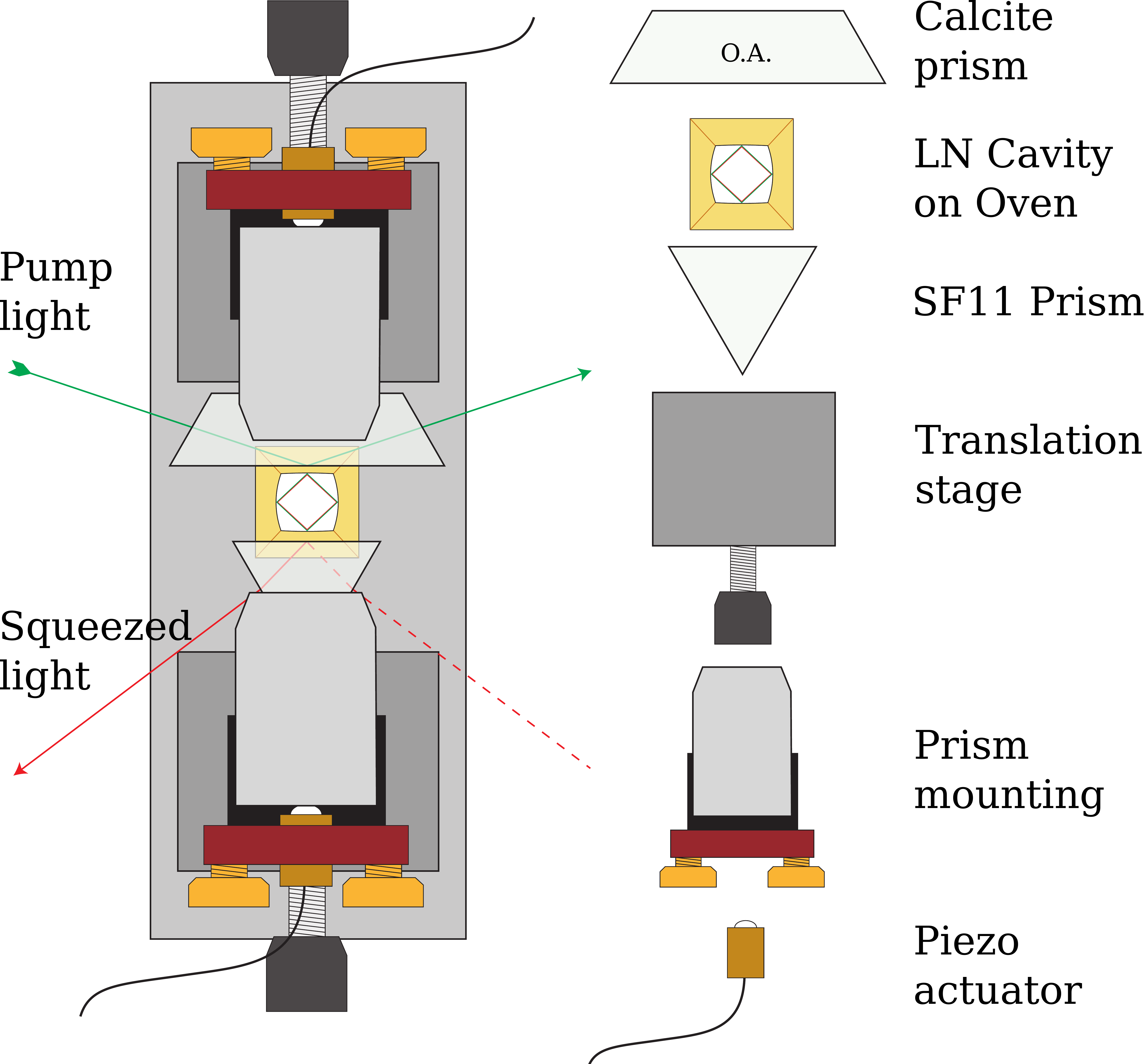}
\par\end{centering}
\caption{Schematic of the set-up to prism-couple the resonator. The LN cavity is mounted on a brass oven that is stabilized to the phase-matching temperature. Two prisms, a green calcite prism to couple the pump and a fused silica prism to couple the sub-harmonic, are mounted on positioning stages and can be brought into the evanescent field of the cavity. Each positioning stack consists of a manual translation stage for coarse positioning, a prism mount for angular adjustments and a prism holder that integrates a piezo-driven flexure for fine positioning.}
\label{fig:alignment}
\end{figure}

The prisms are mounted in a custom-designed holder and the distance between each prism and the resonator is actuated using a combination of piezoelectric elements for fine control and manual translation screws for coarse adjustment. Figure~\ref{fig:alignment} illustrates the design of the prisms and positioning apparatus. The angle of the prisms can also be adjusted by high-stability kinematic mounts. The alignment is done by illuminating the coupling interface with a narrow-band incoherent light source and monitoring the ``Newton's rings'' interference pattern that forms between the prism and resonator surfaces.

Independent control of the prism distances provides significant flexibility in tuning the linewidths of the pump and sub-harmonic resonances. Figure~\ref{fig:FinesseGreenRedThreePrism} shows the regions (shaded in yellow) of linewidths that can be accessed for different prism configurations. The regions are calculated using the anisotropic Fresnel equations for transmission through an evanescent gap (derivation provided in the appendix) along with a sampling of linewidth values measured experimentally. By displacing the coupling prism (either the red calcite prism or the SF11 prism, depending on configuration), the coupling rate for the sub-harmonic is continuously variable from the absorption-limited linewidth of $9$~MHz to nearly the cavity free-spectral range of 18.7 GHz (not shown in the plot). Similarly, the coupling rate of the pump can be varied from 20 MHz (a Q-factor of $2.8\times 10^7$) to roughly 700 MHz by displacing the coupling prism (either the SF11 prism or the green calcite prism). The maximum coupling rate is limited by the parallelism of the prism relative to the resonator surface, an effect that is more pronounced for the pump due to the shorter wavelength.

\section{Self-locking and double resonance}

\begin{figure}
\begin{centering}
\includegraphics[width=0.7\columnwidth]{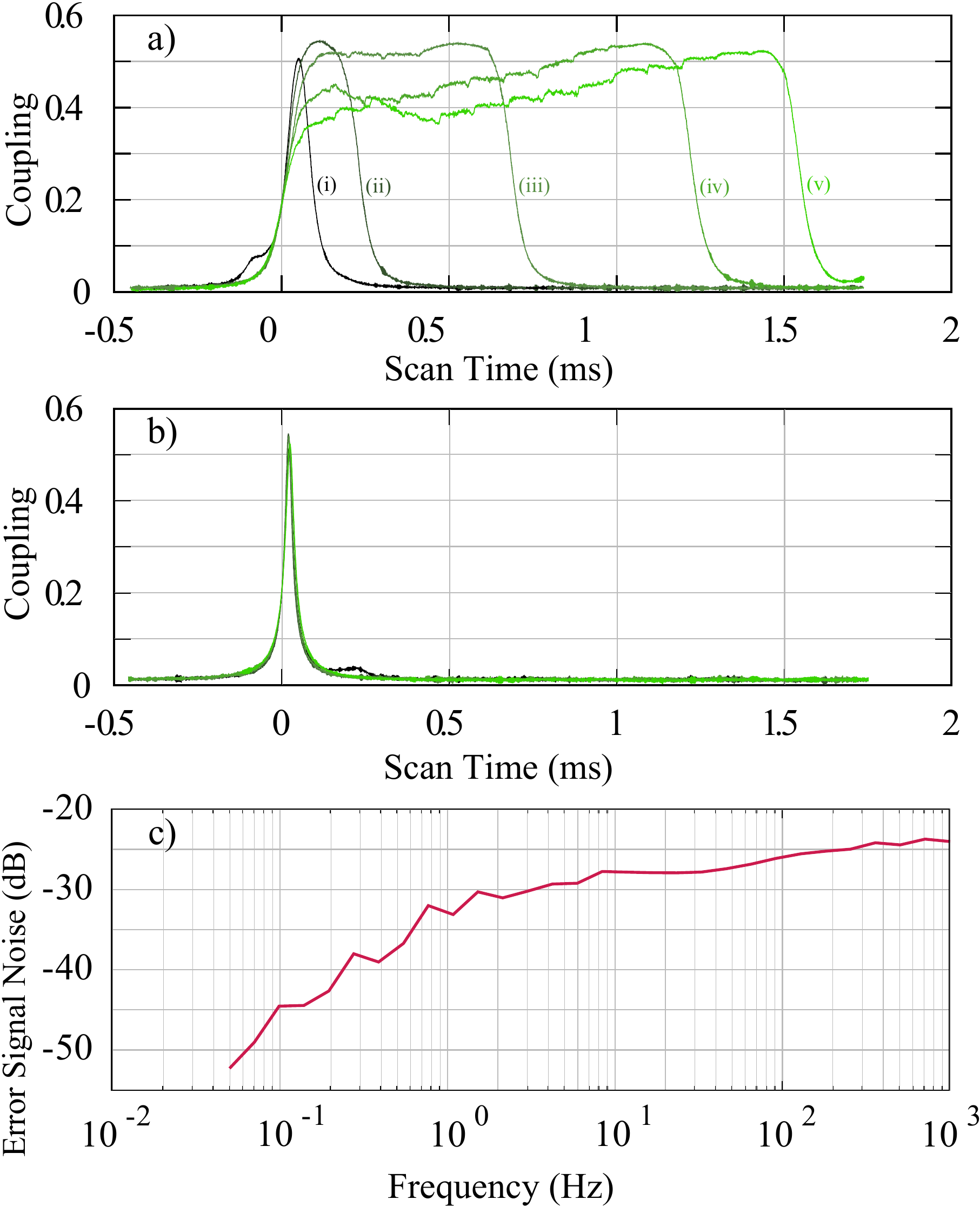}
\par\end{centering}
\caption{\label{fig:self-locking} Demonstration of self-locking of the pump field to a cavity resonance. The resonance behavior of the pump is shown for a scan with increasing (a) and decreasing (b) speed at different pump powers: 1.07 mW (i), 1.18 mW (ii), 1.21 mW (iii), 1.24 mW (iv) and 1.3 mW (v). The different peaks have been realigned to a common origin for more clarity. (c) A network analyzer is used to modulate the laser frequency while monitoring the amplitude noise of the pump light reflected from the cavity while self-locked. The 3 dB roll-off of the noise reduction occurs at around 40 Hz for the pump powers that were typically used.}
\end{figure}

The voltage tuning that is used to scan the cavity resonances for alignment purposes could be used to lock the cavity on resonance. At the pump field intensities that are required to obtain squeezing, however, any externally applied voltage amplifies the strong photo-refractive effects already present. These effects, resulting from optically-induced charge mobility \cite{bryan1985magnesium,volk1994optical}, lead to strong instabilities. Although detrimental to active voltage control, these effects can be useful as they enable passive locking of the nonlinear cavity. Within a particular range of pump powers, the cavity will self-lock~\cite{SelfLocking,Chow:05} near resonance, as shown in Fig.~\ref{fig:self-locking}.

The self-locking is a result of feedback to the cavity resonance frequency as a result of photo-thermal and photo-refractive effects. While scanning the laser frequency, the effect can be observed as a broadening of the linewidth when approaching resonance from one side, as shown in Fig.~\ref{fig:self-locking}(a), and as a narrowing of the linewidth when approaching from the other, as shown in Fig.~\ref{fig:self-locking}(b). We observe the effect only for the pump field, which is absorbed more strongly than the sub-harmonic, and only at relatively high input powers.

We achieve self-locking by approaching the pump resonance from the side that exhibits negative feedback. The bandwidth and self-locking point depend on the linewidth and on the power of the pump field. We find that self-locking effect with a bandwidth of roughly 40 Hz, shown in Fig.~\ref{fig:self-locking}(c), occurs under the conditions used to obtain squeezing.

\begin{figure}
\begin{centering}
\includegraphics[width=0.8\columnwidth]{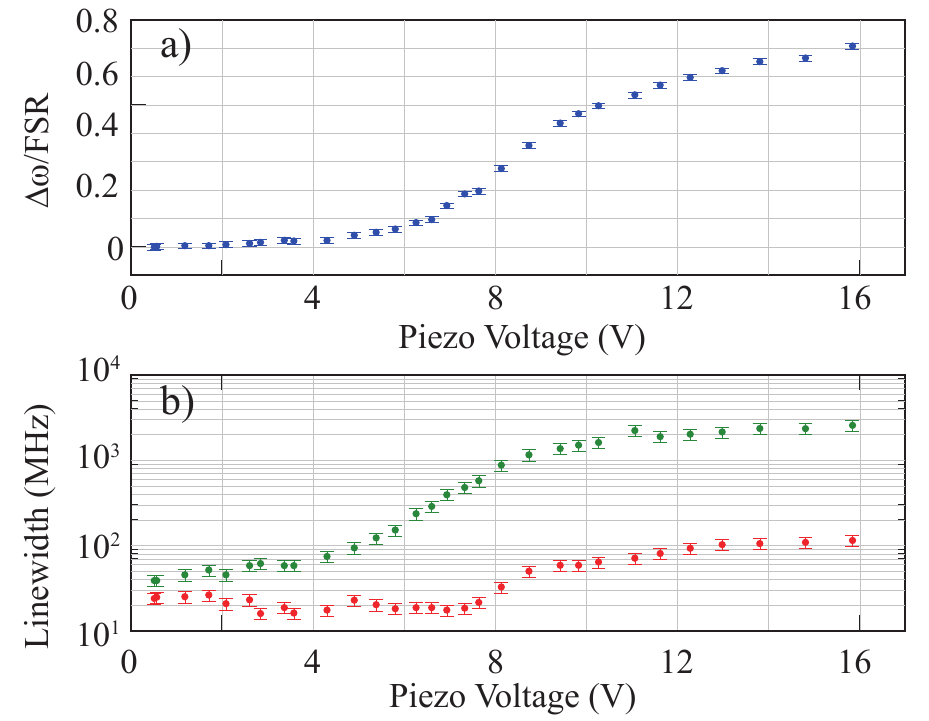}
\par\end{centering}
\caption{\label{fig:redphasecontrol} (a) The relative frequency separation between the pump and sub-harmonic resonances as a function of the piezo tuning voltage for the green calcite prism. (b) The linewidths for the pump (green) and sub-harmonic (red) for the same tuning range. Ideally, the linewidth for the sub-harmonic would be invariant as the distance of the green calcite prism is changed; however, imperfections in the alignment result in some leakage to the green calcite prism.}
\end{figure}

We employ a combination of techniques to tune the resonator to achieve double resonance for both pump and sub-harmonic modes. The temperature of the crystal, initially set near the center of the phase-matching point for 1064/532 nm light (60 $^\circ$C), is tuned to bring the resonances of the pump and sub-harmonic to within roughly 1 GHz of each other. This is accomplished by monitoring the reflections of the pump and a reference beam at 1064 nm while voltage-tuning the resonator. The pump power is kept at a level that is low enough to avoid photo-refractive effects. Once near double resonance, the applied voltage is reduced to zero and the power of the pump is increased until self-locking is attained. Double resonance is then achieved by adjusting the distance of the green calcite coupling prism, similar to the tuning method used in \cite{schunk2015interfacing}. For the ordinarily polarized sub-harmonic, the refractive index of the green calcite prism is low enough that no light will be coupled out by this prism. The presence of the green calcite within the evanescent field will, however, alter the phase of the total internal reflection so the prism can be used to control the resonant frequency of the sub-harmonic, providing faster feedback than would be possible using only temperature tuning. Figure~\ref{fig:redphasecontrol}(a) demonstrates that the relative phase between the pump and sub-harmonic can be tuned by nearly a free-spectral range.

The technique allows the possibility of achieving double-resonance at the optimal phase-matching temperature, although we found that the critical temperature range for phase-matching was roughly 1 $^\circ$C and that fine-tuning within this range did not yield substantial improvements. A downside to this tuning method is that the linewidth of the pump resonance is also affected by the prism distance, as is shown in Fig.~\ref{fig:redphasecontrol}(b), resulting in a coupling of tuning parameters that would ideally be independent. Additionally, imperfections in the coupling, possibly a result of a misalignment between the optical axis of the resonator and that of the prism, lead to some loss of the sub-harmonic through the green calcite prism. Both of these problems could be mitigated by the introduction of a third prism with a sufficiently low index of refraction that neither of the fields can be out-coupled, as was used in \cite{schunk2015interfacing}.

\section{Results}

\begin{figure}
\begin{centering}
\includegraphics[width=0.8\columnwidth]{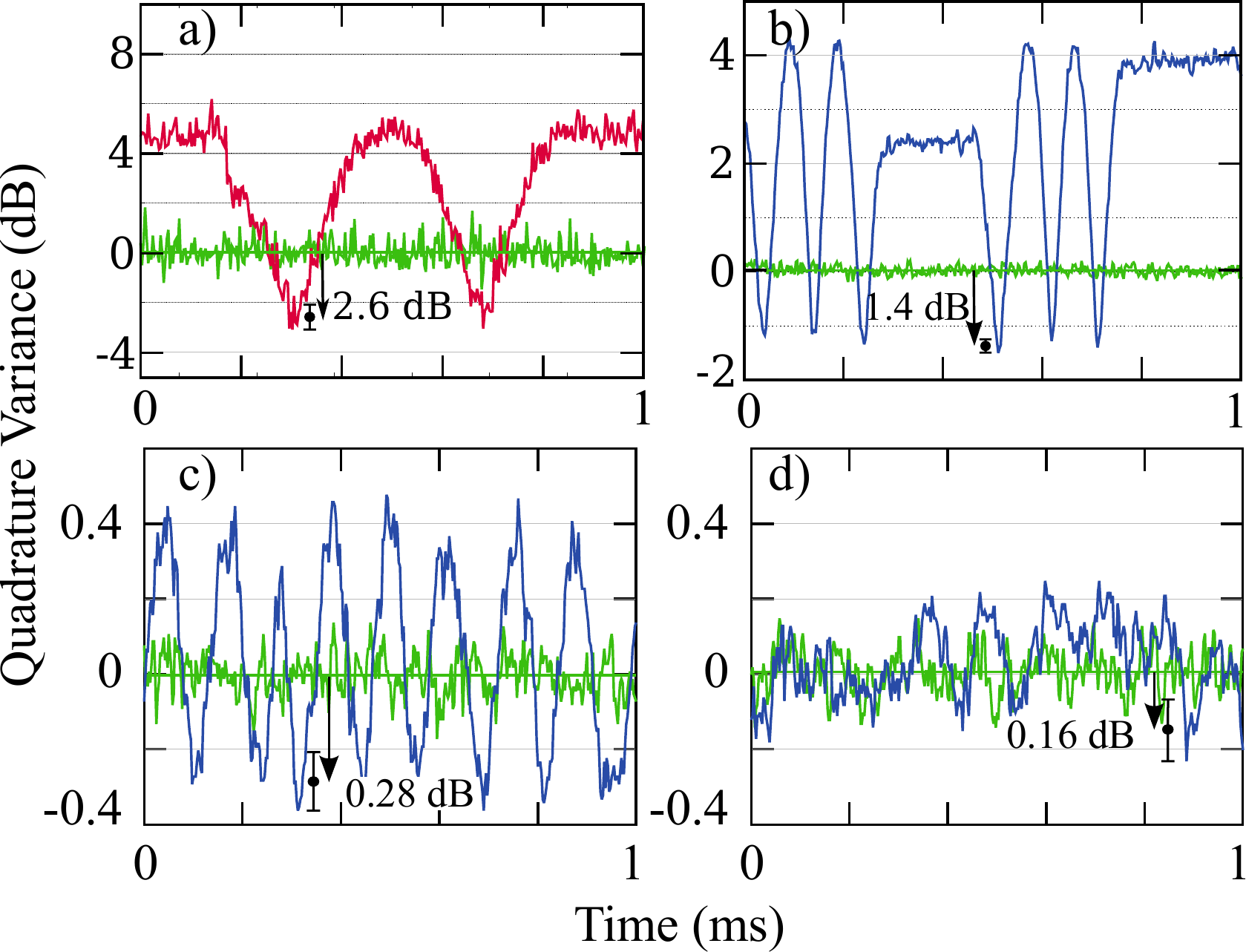}
\par\end{centering}
\caption{\label{fig:squeezing} Observation of squeezing and anti-squeezing by scanning the local oscillator phase. For (a), the red and green calcite prisms are used to independently couple the pump and squeezed field, the local oscillator phase is scanned at 1~Hz and the measurement is performed at a sideband frequency of 5~MHz. For (b-d) the green calcite prism was used in conjunction with the SF11 prism, the local oscillator was scanned at $~3$~Hz and the measurement sideband frequency was (a) 3~MHz, (b) 100~MHz and (c) 150~MHz. The flat regions in (a) and (b) are due to a saturation of the amplifier that provides the piezo scan for the local oscillator phase.}
\end{figure}

We perform a homodyne measurement of the squeezed vacuum field by interfering the out-coupled sub-harmonic with a bright local oscillator. The local oscillator is aligned to a reference beam that has been mode-matched to optimally couple into the resonator through the sub-harmonic coupler prism. The reference beam is then blocked to measure vacuum squeezing. Figure~\ref{fig:squeezing}(a) shows the squeezing that was obtained from the resonator. We observed a maximum of $2.6\pm0.5$~dB ($4.7$~dB corrected) of squeezing below the shot noise level at a sideband frequency of 5~MHz. The green calcite prism and red calcite prism were used to couple the pump and squeezed field, respectively. The finesse for the sub-harmonic is around 100, the finesse for the pump is around 200 and the input power for the pump is around $10$~mW. All three of these values were varied slightly until the resonator was stably self-locked and doubly resonant. The correction to the observed squeezing takes into account the dark noise floor, which is 18~dB below the shot noise level for each measurement, 98\% quantum efficiency of the detectors, 85\% visibility of the homodyne detection and 1.6\% loss due to the various optics after the coupling prism. No correction is made for intra-cavity losses due to scattering, absorption or leakage through the pump coupling prism.

Measurements of squeezing were also taken using a combination of the green calcite prism to couple the pump and the SF11 prism to couple the squeezed field. For this configuration, we observed a maximum of $1.4\pm0.1$~dB ($2.6$~dB corrected) of squeezing at a sideband frequency of 3~MHz with an input pump power of around $200$~mW.  The corrected squeezing value for this measurement takes into account a homodyne visibility of 84\% and an additional loss of 7.4\% due to the squeezed field leaving the SF11 prism at normal incidence. In this configuration, the proximity of the SF11 prism to the resonator leads to a low finesse for the pump and requires pump coupling prism to be in close proximity to the resonator to achieve impedance matching. The pump coupling prism then introduces some losses for the squeezed field, either due to scattering or imperfect alignment of the optic axis of the prism, reducing the observed squeezing.

The small size of the resonator also affords a wide squeezing bandwidth. The free spectral range (FSR) of the cavity is $18.7\pm0.5$~GHz and the finesse of $\sim 100$ for the sub-harmonic provides a squeezing bandwidth of roughly 190~MHz. Figures~\ref{fig:squeezing}(b) and  \ref{fig:squeezing}(c) demonstrate $0.28\pm0.08$~dB (0.8~dB corrected) of shot-noise squeezing at 100~MHz and $0.16\pm0.08$~dB (0.5~dB corrected) at 150~MHz with the SF11 output coupler prism set-up. The detector used for these high-frequency measurements has a bandwidth of 3~GHz but an electronic noise floor that is only 4~dB below shot noise, resulting in higher corrections.

Although the amount of squeezing observed is modest, there are a number of known inefficiencies in our system that can be redressed. The relatively low visibility of the homodyne detection is primarily due to the mode shape being perturbed after excessive polishing  of the trial system, potential misalignment of the extraordinary axis of the resonator with the vertical and astigmatism due to a non-optimal shape of the resonator surfaces. These problems could be easily improved in further iterations of the experiment.

\section{Conclusions}

Monolithic resonators offer significant promise for the development of compact and stable devices for producing quantum states of light. We present the first demonstration of vacuum squeezing from a monolithic resonator design that is based entirely on total internal reflection. The approach combines a simple coating-free fabrication process with the flexibility of independently tunable coupling rates for the pump and squeezed fields and the passive stability obtained by self-locking of the resonator to the pump field.

While the amount of squeezing obtained was a modest 2.6 dB below the shot noise, there is significant room for improvement. The loss introduced through leakage of the squeezed field into the green calcite prism could be reduced by improving the tolerances in the fabrication process. Furthermore, the mirror curvatures could be engineered to maximize the non-linear interaction between the pump and sub-harmonic \cite{paschotta1994nonlinear}. The curvatures used in our resonator were smaller than ideal for ease of coupling and may have changed further due to repeated polishing during testing. The introduction of an evanescent tuning prism that is separate from the pump coupling prism would also simplify the optimization process for achieving double resonance and would allow for a more complete exploration of the parameter space consisting of prism distances, pump power, and resonator temperature. With these  improvements, we see no reason why the system could not achieve the performance of other experiments using bulk crystals in free space \cite{Shnabellithium,0264-9381-29-14-145015,takeno2007observation}.

\section*{Appendix: Evanescent coupling and phase shift}

\begin{figure}
\begin{centering}
\includegraphics[width=0.8\columnwidth]{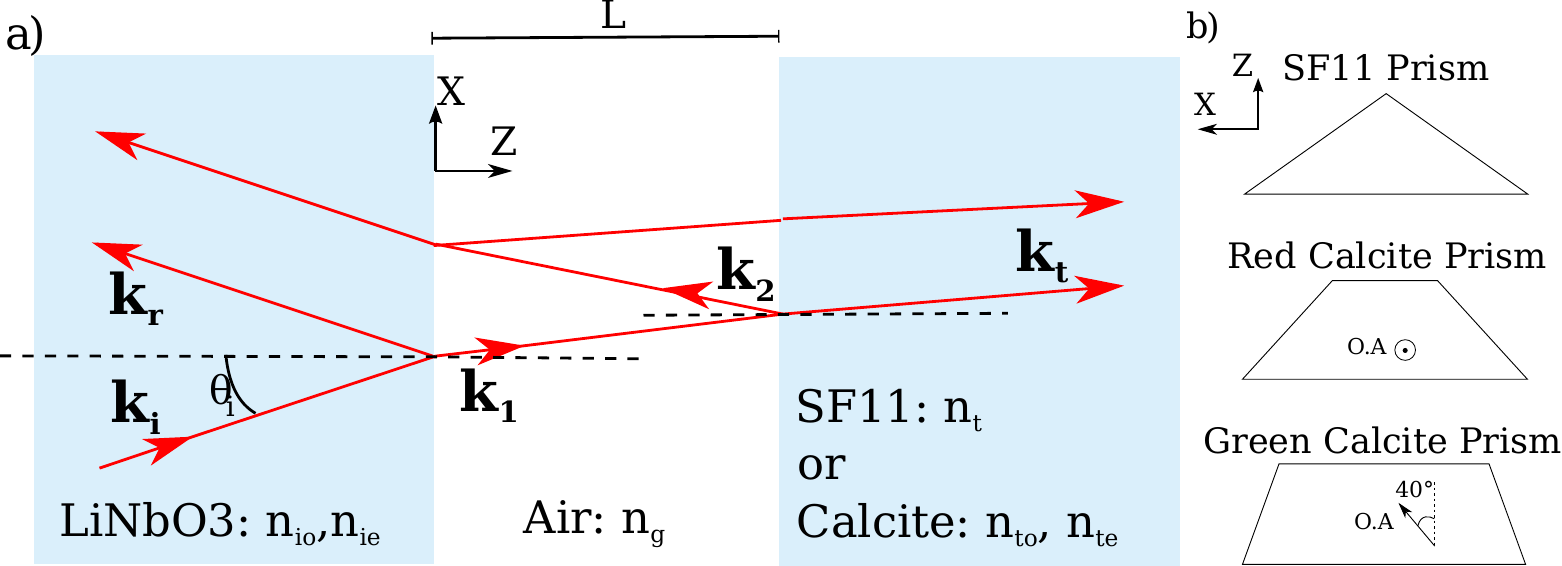}
\par\end{centering}

\caption{(a) Schematic of the coupling media. (b) The three prisms which can be used as third medium.\label{fig:schematic}}
\end{figure}

We calculate the reflection and transmission coefficients for evanescent coupling between the resonator and the coupling prisms. We consider a ray propagating through two interfaces, as is illustrated in Fig.~\ref{fig:schematic}. The first medium, the resonator, has an optical axis that is oriented out of the plane of propagation. The final medium, the coupling prism, may either have an isotropic refractive index, as is the case for the SF11 prism, or may be birefringent with an optical axis that lies in the plane of propagation, as is the case for the green calcite prism,  or in the vertical direction, as the case of the red calcite prism. We assume that the middle medium is air and approximate the refractive index as $n_{g} \approx 1$.

We start with an incident plane wave of wavevector $\mathbf{k}_{i}=k_{ix}\mathbf{e}_{x}+k_{iz}\mathbf{e}_{z}=n_{i}k_{0}\sin(\theta_{i})\mathbf{e}_{x}+n_{i}k_{0}\cos(\theta_{i})\mathbf{e}_{z}$, where $n_{i}$ is the index of the first medium (the ordinary or extraordinary index of the resonator depending on the polarization of the incident field), $k_{0}=\omega/c$ is the vacuum wavevector, with $\omega$ the frequency and $c$ the speed of the light. The directions of the unit vectors $\mathbf{e}_{\lbrace x,y,z \rbrace}$ are indicated in Fig.~\ref{fig:schematic} and $\theta_{i}$ is the angle between $\mathbf{k}_{i}$ and the normal to the first surface.

Using the conservation of $\mathbf{k}_{\parallel}$ and the fact that $|\mathbf{k}_{1}|=|\mathbf{k}_{2}|=n_{g}k_{0}$, $|\mathbf{k}_{r}|=n_{i}k_{0}$, and $|\mathbf{k}_{t}|=n_{t}k_{0}$ we can infer the other wave vectors,
\begin{align}
  \mathbf{k}_{r}&=k_{ix}\mathbf{e}_{x}-k_{iz}\mathbf{e}_{z}, \\
  \mathbf{k}_{1}&=k_{ix}\mathbf{e}_{x}+i\alpha\mathbf{e}_{z}, \\
  \mathbf{k}_{2}&=k_{ix}\mathbf{e}_{x}-i\alpha\mathbf{e}_{z}, \\
  \mathbf{k}_{t}&=k_{ix}\mathbf{e}_{x}+{k}_{tz}\mathbf{e}_{z},
\end{align}
where ${k}_{tz}=((n_{t}k_{0})^{2}-(k_{ix})^{2})^{1/2}$ can be real or purely imaginary, with $n_{t}$ being the refractive index of the last prism, and $\alpha$ is a real number such that $\alpha=\sqrt{\left(k_{ix}\right)^{2}-\left(n_{g}k_{0}\right)^{2}}$.

We consider six cases: \textit{s}-polarized (pump) light and \textit{p}-polarized (sub-harmonic) light, each being coupled with either the SF11 (isotropic) prism or the two  calcite (birefringent) prisms. For the SF11 prism, the index $n_{t}$ is just the index of the prism. For the red calcite prism, in the \textit{s}-polarization case $n_{t}$ is the extraordinary index of the prism $n_{t}=n_{e}$, and  in the the \textit{p}-polarization $n_{t}$ is the ordinary index of the prism $n_{t}=n_{o}$.  For the green calcite prism, in the \textit{s}-polarization case $n_{t}$ is the ordinary index of the prism $n_{t}=n_{o}$ while for the \textit{p}-polarization case we use
\begin{equation}
k_{0}^{2}=\frac{k_{z'}^{2}}{n_{o}^{2}}+\frac{k_{x'}^{2}}{n_{e}^{2}},
\end{equation}
where ($k_{x'},k_{y'},k_{z'})$ are the coordinates of the wave vector in
the reference frame of the anisotropic axis of the prism. Transforming to the reference frame of the problem by applying $k_{x'}=k_{tx}\cos(\theta)-k_{tz}\sin(\theta)$, $k_{y'}=k_{ty}=0$, and $k_{z'}=k_{tz}\cos(\theta)+k_{tx}\sin(\theta)$, with $\theta$ being the angle between the extraordinary axis and $\mathbf{e}_x$, and $k_{tx}=k_{ix}$, we get an equation of second order in $k_{tz}$,
\begin{equation}
k_{tz}=-\frac{k_{ix}n_{p1}^{2}}{\gamma}\pm n_{p1}^{2}{\Delta}^{1/2}\label{eq:ktp-1},
\end{equation}
with $\Delta=\frac{k_{0}^{2}}{n_{p1}^{2}}-k_{ix}^{2}\Big(\frac{1}{n_{p1}^{2}n_{p2}^{2}}-\frac{1}{\gamma^{2}}\Big)$,
$\frac{1}{n_{p1}^{2}}=\frac{\cos^{2}(\theta)}{n_{o}^{2}}+\frac{\sin^{2}(\theta)}{n_{e}^{2}}$,
$\frac{1}{n_{p2}^{2}}=\frac{\cos^{2}(\theta)}{n_{e}^{2}}+\frac{\sin^{2}(\theta)}{n_{o}^{2}}$
and $\frac{1}{\gamma}=\cos(\theta)\sin(\theta)\Big(\frac{1}{n_{o}^{2}}-\frac{1}{n_{e}^{2}}\Big)$.
In our case $\Delta$ is negative because $ \theta=40^{\circ}$, meaning that $k_{tz}$ is complex and we can calculate $n_{t}={\Big(\frac{k_{tz}^{2}+k_{ix}^{2}}{k_{0}^{2}}\Big)}^{1/2}$, which is also complex.

The \textit{s}-polarization case is straightforward, with all the electric fields polarized out of the plane of propagation, along $\mathbf{e}_{y}$. Following the calculation of the Fresnel reflection and transmission coefficients, we use the continuity of the field at both interfaces. The tangential direction of the magnetic field is also continuous, making the field $\partial E_{y}/\partial z$ continuous at both interfaces. We obtain four equations with four unknown variables and, by solving this system, we can get the transmitted and reflected coefficients of the whole system as a function of the distance $L$ between the prism and the resonator:
\begin{align}
t_{s}&=2\Big( \frac{1}{i\alpha}\left[i\alpha\cosh(\alpha L)+k_{tz}\sinh(\alpha L)\right]+\frac{1}{k_{iz}}\left[i\alpha\sinh(\alpha L)+k_{tz}\cosh(\alpha L)\right] \Big)^{-1}, \\
r_{s}&=\frac{\frac{1}{i\alpha}\left[i\alpha+k_{tz}\tanh(\alpha L)\right]-\frac{1}{k_{iz}}\left[i\alpha\tanh(\alpha L)+k_{tz}\right]}{\frac{1}{i\alpha}\left[i\alpha+k_{tz}\tanh(\alpha L)\right]+\frac{1}{k_{iz}}\left[i\alpha\tanh(\alpha L)+k_{tz}\right]}.
\end{align}

For the case of \textit{p}-polarisation, we use the continuity of the magnetic fields $\mathbf{H}=H\mathbf{e}_{y}$ in all media. This gives
\begin{align}
H_{0}+r_{s}H_{0} & = H_{10}+H_{20},\\
H_{10}e^{-\alpha L} + H_{20} e^{\alpha L} & = t_{H}H_{0}.
\end{align}
We use the Maxwell-Amp\`ere equation $\nabla\times\mathbf{H}=\frac{\partial H_{y}}{\partial z}\mathbf{e}_{x}-\frac{\partial H_{y}}{\partial x}\mathbf{e}_{z}=-i\omega\mathbf{D}=-i\omega(\boldsymbol{\bar{\epsilon}}\mathbf{E})$, where $\boldsymbol{\bar{\epsilon}}$ is the permitivity tensor of the material, to get the other two equations. The transverse component of the electric field is continuous, so the field
\begin{equation}
\left(\boldsymbol{\bar{\epsilon}}^{-1}\left(\frac{\partial H_{y}}{\partial z}\mathbf{e}_{x}-\frac{\partial H_{y}}{\partial x}\mathbf{e}_{z}\right)\right)_{\mathbf{e}_{x}}=iH_{0}\left(\boldsymbol{\bar{\epsilon}}^{-1}\left(k_{z}\mathbf{e}_{x}-k_{x}\mathbf{e}_{z}\right)\right)_{\mathbf{e}_{x}}
\end{equation}
is also continuous at the interfaces. For the resonator this corresponds to $iH_{0}k_{iz}/n_{io}^{2}$ for the incident field where $H_{0}$ is the amplitude of the field and $-ir_{H}H_{0}k_{iz}/n_{io}^{2}$ for the reflected beam, where $r_{H}$ is the coefficient of reflection of the system. In the evanescent gap, it corresponds to $iH_{1}k_{1z}/n_{g}^{2}$ and $-iH_{2}k_{1z}/n_{g}^{2}$, with $H_{1}$ and $H_{2}$ being the amplitude of the fields. 

\begin{figure}
\begin{centering}
\includegraphics[width=0.7\columnwidth]{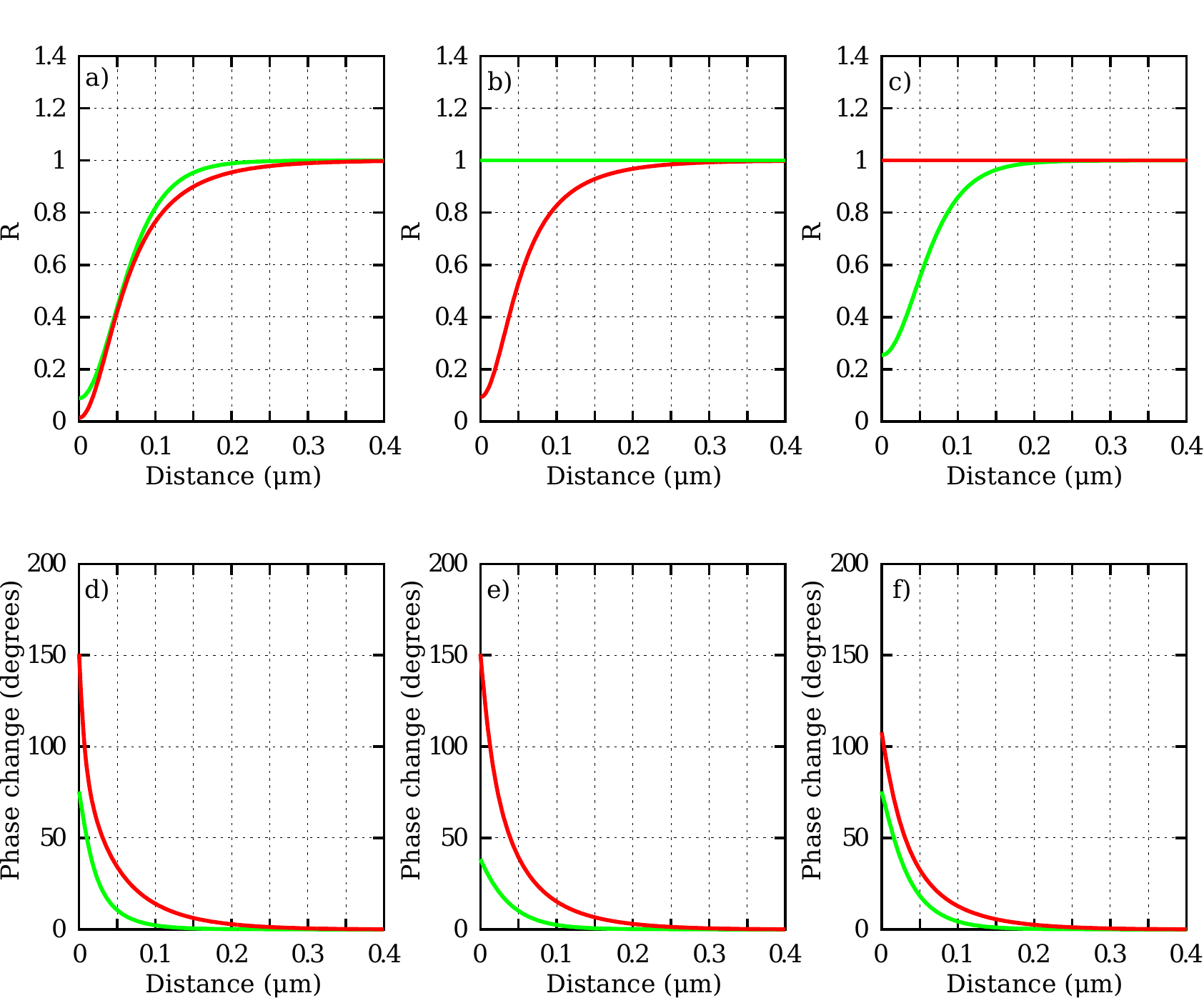}
\par\end{centering}
\caption{\label{fig:fresnel} Reflectivity $R$ versus distance $L$ between the prism and the resonator for 1064 nm (red) and 532 nm (green) for the SF11 prism (a), the red calcite (b) and the green calcite prism (c), and phase shift versus distance due to the reflection for the SF11 prism (d), the red calcite (e) and the green calcite prism (f).}
\end{figure}

For the case of the isotropic SF11 prism or the birefringent red calcite, we get $it_{H}H_{0}k_{tz}/n_{t}^{2}$ with $t_{H}$ the transmission coefficient of the system. The equations are analogous to the previous case with the replacements $\mathbf{E}\to\mathbf{H}$, $t_{s}\to t_{H}$, $r_{s}\to r_{H}$, and $k_{z}\to k_{z}/n^{2}$. For each medium we get
\begin{equation}
r_{H_{SF11}}=\frac{\frac{n_{g}^{2}}{i\alpha}\left[\frac{i\alpha}{n_{g}^{2}}+\frac{k_{tz}}{n_{t}^{2}}\tanh(\alpha L)\right]-\frac{n_{io}^{2}}{k_{iz}}\left[\frac{i\alpha}{n_{g}^{2}}\tanh(\alpha L)+\frac{k_{tz}}{n_{t}^{2}}\right]}{\frac{n_{g}^{2}}{i\alpha}\left[\frac{i\alpha}{n_{g}^{2}}+\frac{k_{tz}}{n_{t}^{2}}\tanh(\alpha L)\right]+\frac{n_{io}^{2}}{k_{iz}}\left[\frac{i\alpha}{n_{g}^{2}}\tanh(\alpha L)+\frac{k_{tz}}{n_{t}^{2}}\right]}.
\end{equation}
For the birefringent green calcite prism we need to apply a rotation $\theta$ around $\mathbf{e}_{y}$ to the inverse of the permittivity tensor of the prism to get its value in the reference frame ($\mathbf{e}_{x},\mathbf{e}_{y},\mathbf{e}_{z})$:
\begin{equation}
\epsilon^{-1}_{(\mathbf{e}_{x},\mathbf{e}_{y},\mathbf{e}_{z})}=\left(\begin{array}{ccc}
1/n_{p1}^{2} & 0 & -1/\gamma\\
0 & 1/n_{o}^{2} & 0\\
-1/\gamma & 0 & 1/n_{p2}^{2}
\end{array}\right).
\end{equation}

The continuous field in the prism is: $it_{H}H_{0}\left(\frac{k_{tz}}{n_{p1}^{2}}+\frac{k_{ix}}{\gamma}\right)$. Defining $k_{tp}=\frac{k_{tz}}{n_{p1}^{2}}+\frac{k_{ix}}{\gamma}$ and noting that, following from equation \ref{eq:ktp-1}, $k_{tp}$ is purely
imaginary, we get the equations
\begin{align}
H_{0}\frac{k_{iz}}{n_{io}^{2}}-r_{H}H_{0}\frac{k_{iz}}{n_{io}^{2}} & = H_{1}\frac{k_{1z}}{n_{g}^{2}}-H_{2}\frac{k_{1z}}{n_{g}^{2}},\\
H_{1}\frac{k_{1z}}{n_{g}^{2}}e^{-\alpha L}-H_{2}\frac{k_{1z}}{n_{g}^{2}}e^{\alpha L} & = t_{H}H_{0} k_{tp},
\end{align}
which have solutions
\begin{equation}r_{H_{calcite}}=\frac{\frac{n_{g}^{2}}{i\alpha}\left[\frac{i\alpha}{n_{g}^{2}}+k_{tp}\tanh(\alpha L)\right]-\frac{n_{io}^{2}}{k_{iz}}\left[\frac{i\alpha}{n_{g}^{2}}\tanh(\alpha L)+k_{tp}\right]}{\frac{n_{g}^{2}}{i\alpha}\left[\frac{i\alpha}{n_{g}^{2}}+k_{tp}\tanh(\alpha L)\right]+\frac{n_{io}^{2}}{k_{iz}}\left[\frac{i\alpha}{n_{g}^{2}}\tanh(\alpha L)+k_{tp}\right]}.
\end{equation}
Note that the reflectivity $R=|r_{H}|^{2}=1$ because $k_{tp}$ is purely imaginary. Figure~\ref{fig:fresnel} shows the reflection and transmission coefficients (a-c) and the phase shift on reflection (d-f) for the pump and sub-harmonic fields in both the calcite prisms and the SF11 prism.

\section*{Acknowledgments}
We would like to thank Dr J.\ Chow for useful discussions. This research was funded by the Australian Research Council Centre of Excellence CE110001027. PKL acknowledge supported from Tianjin Qianren and ARC Laureate Fellowship FL150100019.

\end{document}